\documentclass[preprint,showpacs,preprintnumbers,amsmath,amssymb]{revtex4}
\usepackage{graphicx}
\usepackage{dcolumn}
\usepackage{bm}

\begin{document}
\preprint{APS/123-QED}

\title{Atomistic treatment of depolarizing energy and field in ferroelectric nanostructures}

\author{I. Ponomareva, I. I. Naumov, I. Kornev, Huaxiang Fu and L. Bellaiche }
\affiliation{Department of Physics, University of Arkansas, Fayetteville, Arkansas 72701, USA }

\date{\today}
\begin{abstract}

An {\it atomistic} approach allowing an accurate and efficient treatment
of depolarizing energy and field in {\it any} low-dimensional 
ferroelectric structure is developed. Application of this approach 
demonstrates the limits of the widely used continuum model (even) for
simple test cases. Moreover,  implementation of this approach within a
first-principles-based model reveals an unusual phase transition -- from a state exhibiting
a spontaneous polarization to a phase associated with a toroid moment of polarization --
in a ferroelectric nanodot for a critical value of the depolarizing field.

\end{abstract}

\maketitle
\newpage

\narrowtext

\maketitle

\marginparwidth 2.7in
\marginparsep 0.5in

Ferroelectric nanostructures (FEN) 
are of increasing technological and fundamental
interest because of the need in miniaturization of devices, as well as, the
appearance of new phenomena (see, e.g., Ref. 
\cite{scott,fong,junquera,kornev,fu,naumov} and references therein).
Unscreened polarization-induced charges at the surfaces of FEN generate a 
depolarizing field that is responsible for striking properties. Examples  
are the existence of a critical thickness below which no ferroelectricity can appear 
\cite{junquera}, and the observation and prediction of laminar stripe nanodomains 
\cite{fong,kornev} as well as the formation of polarization vortex \cite{fu,naumov}.
Interestingly, and despite its huge importance, we are not aware of any model being able
to {\it exactly} calculate the depolarizing field and energy in {\it any } low-dimensional ferroelectric.
For instance, the widely used continuum model (1) neglects the atomistic 
nature of materials, (2) is technically applicable 
only in the limit of large enough systems and (3) can not  predict 
the depolarizing energy/field in the realistic cases of
inhomogeneously polarized samples. 

In this Letter we i) demonstrate that it is possible to 
derive a scheme allowing the exact  {\it atomistic}  
computation of the depolarizing energy and field in 
any low-dimensional FEN; ii) use this scheme to check the accuracy 
of the continuum model for some simple test cases; iii) report  
an unusual phase transition  
between two different kinds of order 
parameter in a ferroelectric nanodot that is driven by
the depolarizing field.

To calculate the depolarizing energy in low-dimensional ferroelectrics, one
first needs to realize that a system under perfect
open-circuit (OC) electrical boundary conditions exhibits a {\it maximum} 
depolarizing field (if the polarization lies along a non-periodic direction), 
while
ideal short-circuit (SC) electrical boundary conditions leads to a complete screening of charges
at the FEN surfaces that fully annihilates any depolarizing 
field. As a result, the depolarizing energy and field experienced by the 
FEN
should involve a {\it difference} between the dipole-dipole interactions 
associated with these two extreme electrical
boundary conditions.
We shall write energy of the dipole-dipole interaction in {\it any}
system in the form 

\begin{equation}
\mathcal{E}_{dip}^{(D)}= {1 \over 2V} \sum_{ \alpha\beta, ij }
Q_{\alpha\beta,ij}^{(S,D)}p_{\alpha}(\mathbf{r}_{i})p_{\beta}(\mathbf{r}_{j}),
\label{eq:2}
\end{equation}
where $D=3,2,1$ stands 
for a system periodic in 3, 2 and 1 directions, respectively; $D=0$
corresponds to non-periodic systems and
the sum  runs over  the 
atomic sites $i$ and $j$ that differ from each other 
and belong to a {\it supercell} (to be denoted by $S$) 
mimicking the system. Such a supercell is infinitely
repeated along the periodic  directions, if any. 
For instance, thin films 
are modeled by supercells that are repeated in two dimensions
while the direction associated with the growth direction of the film 
is non-periodic. For dots, the 
supercell is not
repeated.
V is the volume  of the supercell, $\bm p(\bm r_i)$  the dipole 
moment at the site $i$,
$\alpha=x,y,z$ denotes the Cartesian components. 
The quantity $Q^{(S,D)}$ depends on both the chosen 
supercell ($S$) and the periodicity
of the system ($D$).

It is straightforward to prove that the elements of the Q matrix 
for systems periodic in three\cite{zhong}, 
two ($x$- and $y$), one ($z$) directions\footnote {To calculate the elements
of the Q-matrix we developed an original approach using {\it periodic} Green's
function - satisfying Laplace equation and {\it analytically} determined.
Details of this approach will be published 
elsewhere. Note that other approaches  allowing the  calculation
of  dipolar interactions 
can be found  in Refs.\cite{rhee,jensen,brodka1,brodka2}.} 
and 
non-periodic systems are given by:  
\begin{eqnarray}
\bullet &&Q_{\alpha\beta,ij}^{(S,3)}=\frac{4\pi}{V}\sum_{\mathbf{G}\neq 0}
 \frac{1}{G^2}exp(-\frac{G^2}{4\lambda^2})G_{\alpha}G_{\beta}
cos(\mathbf{G}\mathbf{r}_{ij})-
\frac{4\lambda^3\delta_{\alpha \beta}\delta_{ij}}{3\sqrt{\pi}},
\nonumber\\
\bullet &&Q_{\alpha\beta,ij}^{(S,2)}=\frac{2\pi}{A}\sum_{\mathbf{G}}
\bigg\{G\cos(\mathbf{G}\cdot
\bm\rho_{ij})\bigg [\frac{1}{\sqrt{4\pi}}\:\Gamma\left(
 -\frac{1}{2},\,\frac{G^{2}}{4\lambda^{2}}
 \right)\delta_{\alpha z}\delta_{\beta z}
 +\frac{1}{G^2}\:\text{erfc}\left(
 \frac{G}{2\lambda}
 \right)G_{\alpha}G_{\beta}\bigg ] 
\nonumber\\
 &&+G\,
\exp(-G\,|z_{ij}|)\bigg[(\frac{G_{\alpha}G_{\beta}}{G^2}
-\delta_{\alpha z}\delta_{\beta z})\cos(\mathbf{G}\cdot\bm \rho_{ij})
\,-\frac{G_{\alpha}\delta_{\beta z}}{G}\sin(\mathbf{G}\cdot\bm \rho_{ij})
\frac{z_{ij}}{|z_{ij}|}\bigg]\bigg\}
-\frac{4\lambda^{3}\,\delta_{\alpha \beta}\delta_{ij}}{3\sqrt{\pi}}, 
\nonumber\\
\bullet &&Q_{\alpha\beta,ij}^{(S,1)}=\frac{2}{a}\sum_{\mathbf{G}}
G^{2}\,\cos(\mathbf{G}\cdot
\mathbf{z}_{ij})\nonumber
\bigg\{K_{0}\big(G\rho_{ij}\big) \:
\delta_{\alpha z}\delta_{\beta z}
\nonumber
+\frac{\delta_{\alpha x}\delta_{\beta x}
+\delta_{\alpha y}\delta_{\beta y}}{G\,\rho_{ij}}
K_{1}\big(G\rho_{ij}\big)\nonumber\\
&&-\frac{1}{\rho_{ij}^2}K_{2}\big(G \rho_{ij}\big)\,
\rho_{\alpha,ij}\rho_{\beta,ij}\bigg\}\nonumber
-\frac{2}{a}\sum_{\mathbf{G}}\:
G\,\sin(\mathbf{G}\cdot \mathbf{z}_{ij})
\:K_{1}\big(G\rho_{ij}\big)\,\rho_{ij}^{-1}\nonumber
\times G_{\alpha}
\rho_{\beta,ij}+
\nonumber\\
&&+\frac{1}{a^{3}}\big(\delta_{\alpha x}\delta_{\beta x}+
\delta_{\alpha y}\delta_{\beta y}-
2\delta_{\alpha z}\delta_{\beta z}\big)\sum_{n=-\infty}^{\infty\quad\prime}
\bigg|n+\frac{z_{ij}}{a}\bigg|^{-3},
\nonumber\\
\bullet &&Q_{\alpha\beta,ij}^{(S,0)}={\delta_{\alpha,\beta}\over r_{ij}^3}-
{3r_{\alpha,ij}r_{\beta,ij}\over r_{ij}^5}
\label{eq:6}
\end{eqnarray}
where 
$\mathbf{G}$ are the reciprocal lattice vectors associated with 
the $S$ supercell, 
$\lambda$ is the Ewald parameter\cite{kittel} (which is assumed to be 
large enough), $\delta_{ij}$ is the 
Kronicker symbol, $A$ is the supercell area,  
$\bm \rho_{ij}$ and $z_{ij}$ are the projections of $\bm r_{ij}$ (vector
connecting atomic sites $i$ and $j$) on the
$\{$$x$,$y$$\}$ plane and $z$-axis, respectively.
$\Gamma$ is the incomplete Gamma function 
  and $erfc$ is  the complementary error function,
  $a$ is the supercell length in $z$-direction; 
$K_{n}$ are the modified Bessel's functions.
 (Note, that contributions from $z_{ij}=0$ in $Q_{\alpha\beta,ij}^{(S,2)}$ 
 and contributions from
$\rho_{ij}=0$ in $Q_{\alpha\beta,ij}^{(S,1)}$ should be excluded, 
and that the prime in 
the right side of $Q_{\alpha\beta,ij}^{(S,1)}$
indicates that the term  
$n=0$ has to  be
excluded when  $i=j$.

The dipolar
interactions described by equations (\ref{eq:2})
and (\ref{eq:6}) correspond to {\it ideal OC conditions}
since no charge screening is taken into account in their derivation. The
next question to be addressed is what is the dipole-dipole energy in FEN
under {\it perfect SC conditions}.
Such energy is simply the one described by $D=3$ Q-matrix since infinitely
extended (i.e, bulk) systems 
do not experience any macroscopic depolarizing field\cite{meyer}. 
One can also be
convinced in the correctness of the above statement 
by applying the so-called image method that produces polarization pattern 
identical to the one in the bulk from the polarization pattern in the 
nanostructure.

The $\mathcal{E}_{dep}^{(D)}$ (maximum) depolarizing energy per volume in 
any FEN can now easily be calculated as the difference
in dipole-dipole energies between perfect OC and SC conditions, that is:
\begin{equation}
\mathcal{E}_{dep}^{(D)}=\mathcal{E}_{dip}^{(D)}-\mathcal{E}_{dip}^{(3)}=
\frac{1}{2V}\sum_{ \alpha\beta, ij }
[Q_{\alpha\beta,ij}^{(S,D)}-Q_{\alpha\beta,ij}^{(S,3)}]p_{\alpha}(\mathbf{r}_{i})p_{\beta}(\mathbf{r}_{j})
\label{eq:10}
\end{equation} 
where the sum over $i$ and $j$ run over the sites of the chosen supercell 
of the FEN, and 
where $Q_{\alpha\beta,ij}^{(S,D)}$ are given by Eq.(\ref{eq:6}).
Equation (\ref{eq:10})  is, to the best of our knowledge, 
the first proposed form allowing
an atomistic and exact computation of depolarizing energy in {\it any} 
FEN with {\it any} dipole distribution. 
(Note  
that such form can also be applied to calculate demagnetization energy in low-dimensional {\it magnetic} systems).

We first apply our approach to compute $\mathcal{E}_{dep}^{(D)}$ in some
{\it test} cases.
Here, we limit ourselves to systems adopting a simple cubic structure in which each 
atomic site has a local and equal-in-magnitude dipole moment. 
Note, that all results to be reported here do not depend on the
size used for the periodic direction(s) of the $S$ supercell.

{\it Homogeneous dipole distribution:} let's first investigate FEN
 exhibiting the same local dipole
$\bf{p}$ at any atomic site. We shall present our results in the form of
$\mathcal{E}_{dep}^{(D)}=\gamma\mathcal{E}_{dep}^{(D,cont)}$, where
$\mathcal{E}_{dep}^{(D)}$ is  obtained  from Eq.(\ref{eq:10}), 
while $\mathcal{E}_{dep}^{(D,cont)}$ is the depolarizing
energy predicted by the continuum model. $\gamma$ is thus 
a ``correcting'' coefficient that provides a measure of the 
continuum approach accuracy.

\begin{figure*}
\includegraphics[height=0.85\textheight]{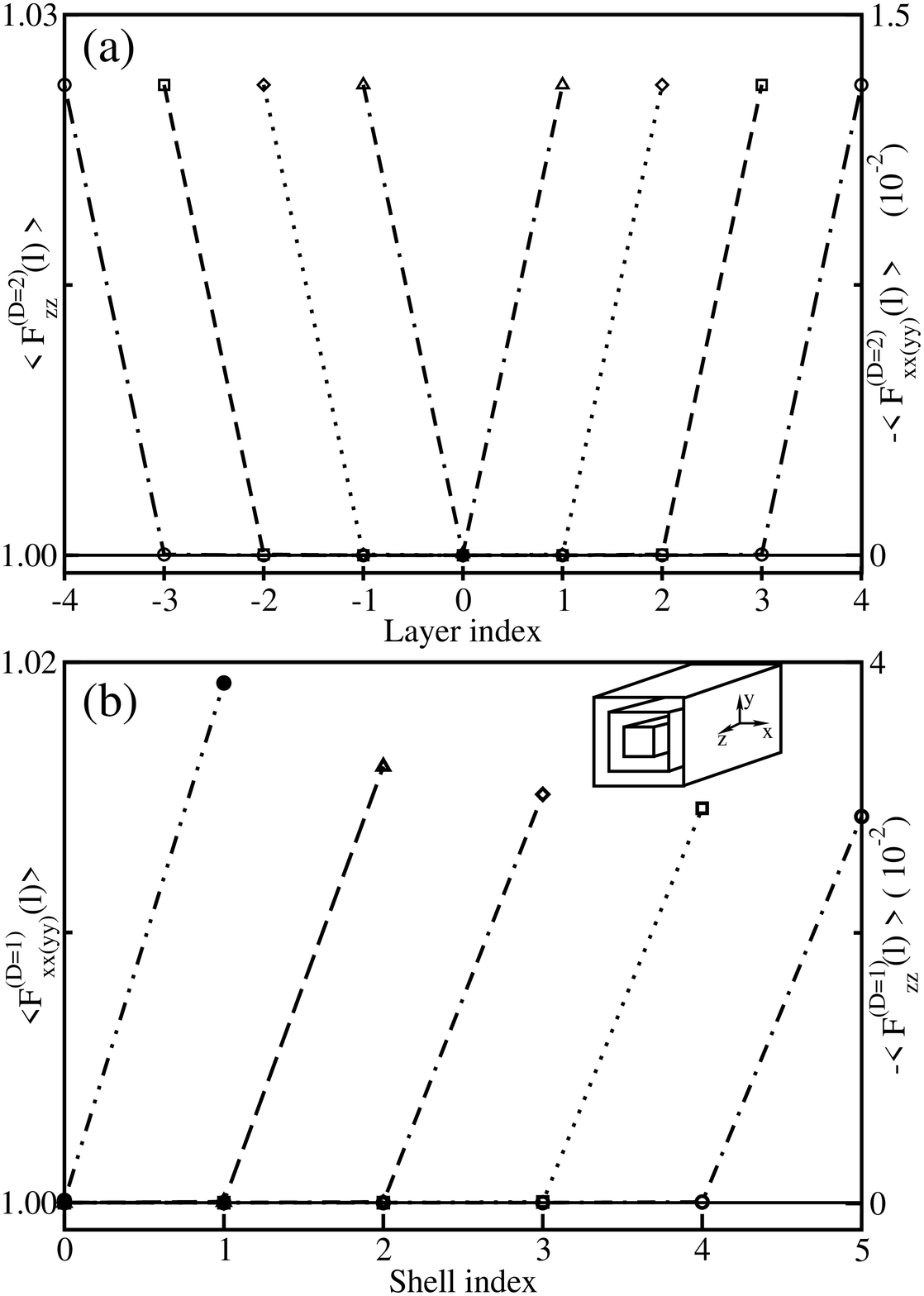}
\caption{\label{fig1}   Depolarizing factors $<F_{xx}^{(D)}(l)>$,
$<F_{yy}^{(D)}(l)>$ and $<F_{zz}^{(D)}(l)>$  obtained with our 
atomistic approach and
normalized to $4\pi$ (that is the prediction of F$_{zz}$ in 
the continuum model) in the case of (001) films (part (a)) and $2\pi$ 
(prediction of F$_{xx}$ in 
the continuum model) in the case of wires periodic along the $z$-axis
  (part (b)) 
as a function of the layer or shell index  $l$. 
The different shells of a wire are shown
in the inset of part (b). The most inner layer or shell is indexed by 0.}
\end{figure*}

For (001) ultra-thin films homogeneously polarized along the out-of-plane 
($z$) direction, the continuum model predicts that
$\mathcal{E}_{dep}^{(2,cont)}=2\pi P^2$, independently of the film thickness, 
where $P$ is the polarization.
On the other hand, the use of Eq~(\ref{eq:10}) results in  
$\gamma=$1.017, 1.010, 1.007 and 1.006
for  ultra-thin films of 3, 5, 7 and 9 layers, respectively.
In other words, our atomistic approach reveals that the depolarizing energy is slightly larger
than the one predicted by the continuum model and increases as the number of film layers decreases. Such  findings are consistent with those 
of Refs.\cite{jensen,vedemenko,draaisma}.
 To understand them, we
rewrite Eq~(\ref{eq:10}) in the case of a {\it  homogeneous} dipole pattern 
as follows:
\begin{equation}
\mathcal{E}_{dep}^{(D)}=
\frac{p_{\alpha}p_{\beta}}{2V}\sum_{ \alpha\beta, j }
F_{\alpha\beta}^{(D)}(j),\;
\begin{text}
with         
\end{text}
F_{\alpha\beta}^{(D)}(j)=\sum_{i} [Q_{\alpha\beta,ij}^{(S,D)}-Q_{\alpha\beta,ij}^{(S,3)}]
\label{eq:10b}
\end{equation} 
Fig.\ref{fig1}a shows the ``depolarizing''  factors $<F_{zz}^{(D=2)}(l)>$ 
averaged over all the $j$ sites belonging to a given (00l) layer 
(that is indexed by $l$) for our  films, 
as a function of a layer position inside the film. Comparison with its continuum predicted (and $l$-independent) 
value of $4\pi$ is also given. 
Fig.\ref{fig1}a  clearly reveals that the deviation of the continuum model 
from our atomistic results is  confined to 
the surface layers - as also found in Ref.\cite{draaisma} - and that this deviation is an underestimation. 
This  explains why we numerically found that the $\mathcal{E}_{dep}^{(2)}$
increases with respect to the continuum prediction as the film becomes thinner,  since the ratio of
surface layers over total layers increases as the film thickness decreases.
Fig.\ref{fig1}a also shows that  for the surface layers 
$<F_{xx}^{(2)}>=<F_{yy}^{(2)}>$ are negative - as discussed in 
Ref.\cite{vedemenko} - 
and that $1+(<F_{xx}^{(2)}>-<F_{zz}^{(2)}>)/4\pi=-0.0393$, 
which is 
in perfect agreement 
with the calculation of the so-called surface anisotropy 
in Ref.\cite{draaisma}. 

We next consider different wires of square cross sections,
that are periodic along the $z$-direction and  homogeneously polarized
 along the $x$-axis. 
According to the continuum approach, such
wires should have a depolarizing
energy $\mathcal{E}_{dep}^{(1,cont)}=\pi P^2$, 
independently of the wire thickness.
We numerically found  via Eq.~(\ref{eq:10}) that 
$\gamma=$1.017, 1.010,  1.007 and 1.006 for wires of 3, 5, 7 and 9 shells, 
respectively (see inset of Fig.\ref{fig1}b for definition of shells).
Like in the  films, the continuum model underestimates the depolarizing
energy and this underestimation becomes larger in magnitude as the 
nanostructure shrinks in size. 
Fig.\ref{fig1}b shows that  the continuum model fails to reproduce 
the averaged depolarizing factor for the
 surface shell but exactly agrees with our atomistic results  
for all the inner shells.  

We now turn our attention to a cubic dot homogeneously
polarized along the z-direction. 
Unlike the previous two cases, 
our atomistic approach gives a 
depolarizing energy that is not only independent on the dot size
but also exactly agrees with the continuum approach 
(that is, $\mathcal{E}_{dep}^{(0)}=\mathcal{E}_{dep}^{(0,cont)}=2\pi P^2/3$).
Such a surprising result is caused by the sum rule \cite{vedemenko}  
for depolarization 
factors ( that is, $F_{xx}+F_{yy}+F_{zz}$ is the constant given 
by the continuum model) that we numerically found  to be valid
in {\it every} layer (shell) of any  system investigated so far 
(i.e. films, wires and dots).  
Moreover,  cubic dots exhibit $x$, $y$ and $z$ directions 
that are symmetrically equivalent 
(which is not the case in wires and thin films). As a result,
 $<F_{xx}^{(D)}(l)>=<F_{yy}^{(D)}(l)>=<F_{zz}^{(D)}(l)>$ in dots.  
Because of the sum rule, each of these factors is  equal to 
the continuum prediction of $4\pi/3$ for any shell and for any size, 
and the continuum model
predicts the right depolarizing energy.

\begin{figure*}
\includegraphics[width=\textwidth]{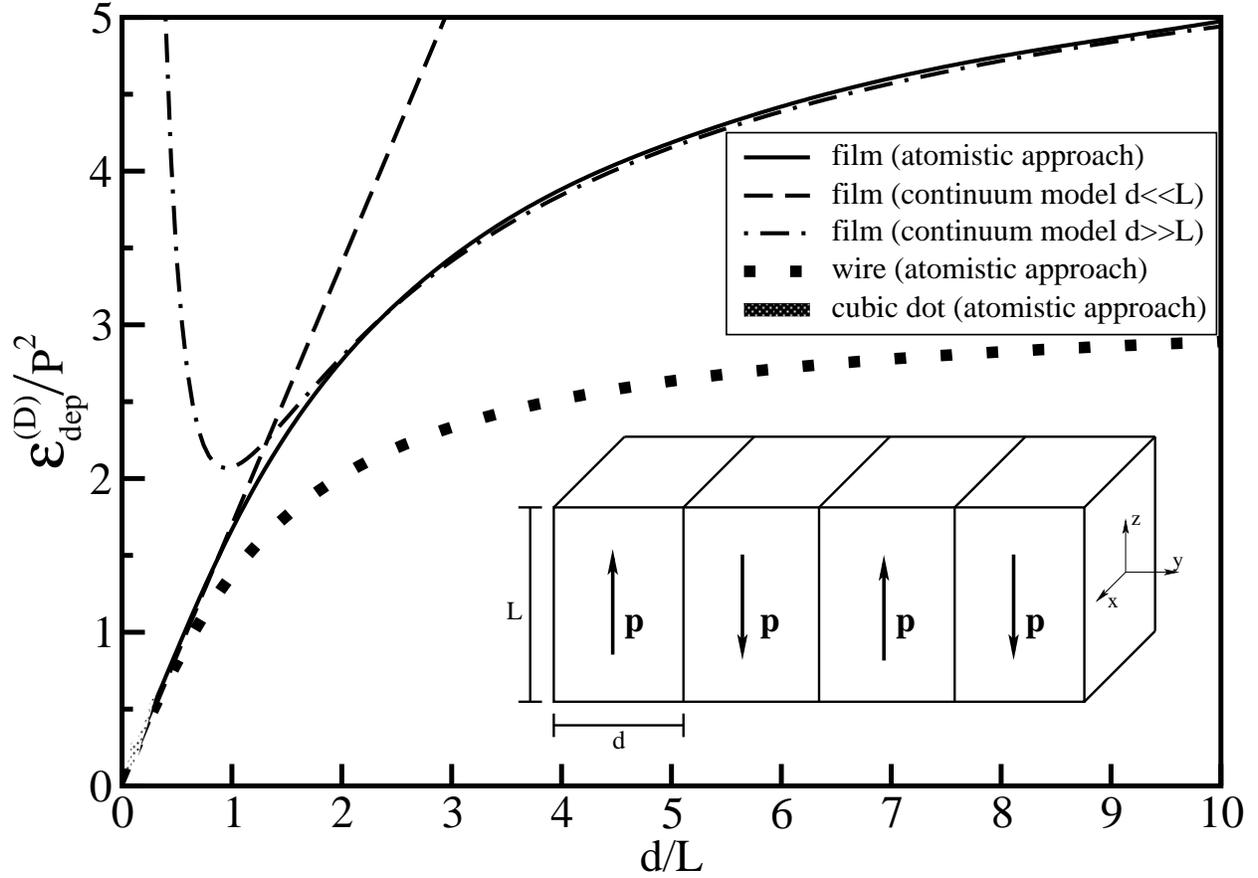}
\caption{\label{fig2}  Normalized  depolarization energy as a function
of $d/L$ for quasi 0D, 1D and 2D systems with stripe domains along with
continuum model prediction for $d<<L$ \cite{kittel2} and 
$d>>L$\cite{kaplan}
 for D=2 systems. The inset shows the schematic representation of the
chosen polarization pattern in (001) films, or wires periodic along $y$ 
or cubic dots.}
\end{figure*}

{\it Inhomogeneous dipole distribution:} let's now investigate 2D, 1D and 0D 
FEN exhibiting
stripe domains with the dipoles assumed to be homogeneous  inside 
each domain (with a $p$ magnitude) and perpendicular to a periodic 
direction, if any
(see the inset of Fig.\ref{fig2}). 
The period of stripe domains is denoted as $d$.
The $\mathcal{E}_{dep}^{(D)}$ energy calculated
 from Eq (\ref{eq:10}) for a   film having a thickness $L=10$
atomic layers  is shown as a function of $d/L$
in Fig.\ref{fig2}, along with depolarization energies 
calculated in 
the continuum approach for the  two limiting cases $d<<L$ 
(i.e, $\mathcal{E}_{dep}^{(2,cont)}=1.7P^2 d/L$ \cite{kittel2,mitsui})
and 
$d>>L$ \cite{kaplan}. 
One can see that these two limiting cases can  reproduce rather well the energy derived from
Eq (\ref{eq:10}) for
$d/L~< 1$ and $d/L~ > 2$, respectively.  
Furthermore we numerically found that our depolarization energy 
can be parametrized as following
\begin{equation}
\mathcal{E}_{dep}^{(D)}=[c_0(1-c_1e^{-d/c_2L})+(n\pi-c_0)(1-e^{-d/c_3L})]P^2
\label{eq:10e}
\end{equation}
where $c_0=2.568$, $c_1=1.024$, $c_2=14.118$, $c_3=1.831$, $n=2$.

Fig.\ref{fig2} also reports  our results 
 for two cases for which we are not aware of any
continuum predictions, namely stripe domains in infinite wires and cubic dots. 
Note that
the stripe direction  is along the wire $y$ periodic 
direction and that the {\it finite} size of 0D systems implies 
that  $d/L$ has a maximum value of 0.5
in cubic dots.  We present here results for a wire of 10 shells 
(10x10 atomic sites for cross section) and for a
 cubic dot of 10 shells (10x10x10 atomic sites) with $d$ ranging from 1 to 5 atomic
layers. 
 Two  features seen in Fig.\ref{fig2} 
are particularly striking. First of all, 
the stripe domains have less depolarizing energy in a wire  than
in a thin film for the same $d/L$, with this difference becoming more pronounced as 
$d/L$ increases. The parameters of equation (\ref{eq:10e}) for the wire are 
$c_0=2.208$, $c_1=1.000$, $c_2=1.196$, $c_3=7.398$ and $n=1$.
Secondly, for the case of 
a cubic dot,  the dependence of depolarizing energy on $d/L$ is
linear and given by $\mathcal{E}_{dep}^{(0)}=2.5P^2d/L$.

We finally take advantage of our formalism to reveal, from realistic
first-principles-based calculations,
properties of a free standing $Pb(Zr_{0.4}Ti_{0.6})O_3$ (PZT) cubic dot of 
48 \AA\, lateral size 
for different  electrical
boundary conditions.  The total energy of the system  used 
in Monte Carlo simulations is:
\begin{equation}
\mathcal{E}_{Heff}({\mathbf{p}(\mathbf{r}_{i})},{\mathbf v_i},\eta,{\sigma _i})+
\beta\sum_i <\mathbf E_{dep}> \mathbf{p}(\mathbf{r}_{i}) 
\label{eq:11}
\end{equation}
where $\mathcal{E}_{Heff}$ is the (first-principles-derived Effective Hamiltonian) energy for PZT
\cite{bellaiche1} which is dependent on the $\mathbf{p}(\mathbf{r}_{i})$ local dipoles
at site $i$ of the dot, 
the $\mathbf v_i$ inhomogeneous strain related variables, the $\eta$ homogeneous 
strain tensor, and on the $\sigma _i$ atomic 
configuration\cite{bellaiche1}. The dipole-dipole interactions
 in this $H_{eff}$ are given 
by the Q-matrix with $D=0$ of Eq.\ref{eq:6}. 
 The second term of equation 
(\ref{eq:11}) mimics a screening of the (maximum) depolarizing field, 
with the magnitude 
of this screening being controlled
by the $\beta$ 
coefficient. $\beta=1$ and $\beta=0$ corresponds to ideal SC and OC 
electrical boundary conditions, respectively, while a value of $\beta$ 
in-between corresponds
to more realistic electrical situation\cite{junquera}.
$<\mathbf E_{dep}>=-(1/N\epsilon_\infty)\sum_j (\partial \mathcal{E}_{dep}^{(D=0)}/\partial
\mathbf p(r_j))$  is the  depolarizing field 
inside the dot, while $N$ and $\epsilon_\infty$ are the total number of sites of the dot and the
dielectric constant of PZT, respectively.
$\mathcal{E}_{dep}^{(D=0)}$ is practically calculated from 
Eq. (\ref{eq:10}).

\begin{figure*}
\includegraphics[width=\textwidth]{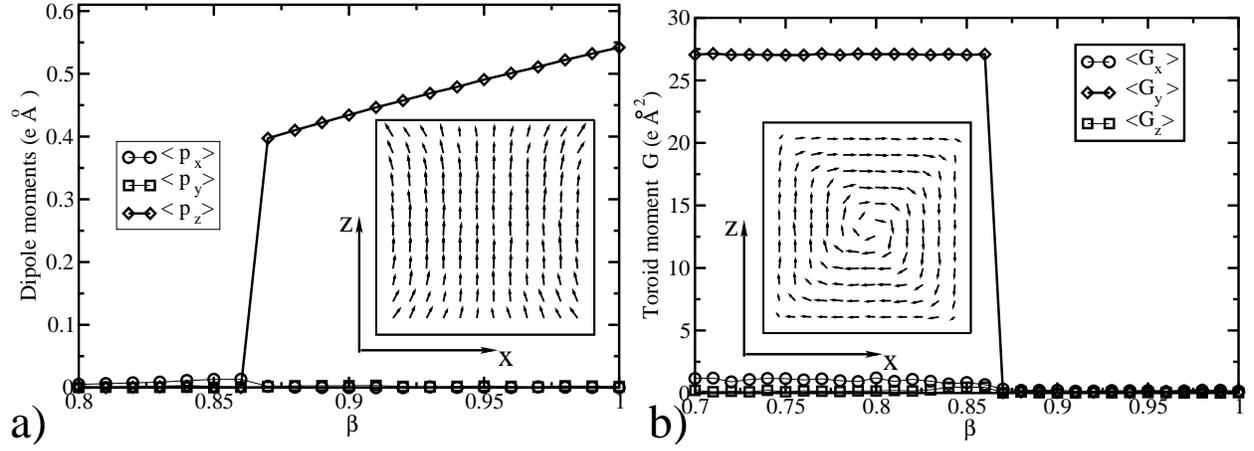}
\caption{\label{fig3}  
a) Dipole moments and b) toroid moment of polarization $G$ in 
a PZT nanodot as a function of the screening coefficient 
$\beta$. Insets of the parts a) and b) show the polarization pattern
for $\beta=1$ (SC conditions) and $\beta=0$ (OC conditions), respectively.
}

\end{figure*}

 Fig.\ref{fig3}a and b show the resulting macroscopic dipole moment 
 and the macroscopic toroid moment of polarization (i.e., the supercell average of the cross product
 between position and dipole moment \cite{naumov}), 
respectively, as a function of
$\beta$. One can clearly see that 
for situations close to
SC, the  dot exhibits a macroscopic {\it polarization}, with a cross section of the
local dipole pattern being given
 in the inset of Fig.\ref{fig3}a. 
On the other hand 
a dot with electrical boundary conditions close to OC 
has a non-vanishing {\it toroid moment}\cite{naumov}, with a cross-section of the corresponding dipole pattern being displayed in the inset
 of Fig.\ref{fig3}b.
 Moreover, Fig.\ref{fig3} clearly reveals that, at a critical value 
of the depolarization field, the system undergoes an unusual
 phase transition between a state characterized by one kind of order parameter 
(toroid moment of polarization) to a state associated with
 another kind of order
parameter (polarization).  
In other words, 
no coexistence between these two order parameters occurs. 

In summary, we have derived an {\it atomistic}, simple, general and efficient approach to 
calculate
the depolarizing energy and field in {\it any} low-dimensional ferroelectric structure.
The application of this method reveals -- and explains --  the limits of  the continuum model, and
also  results in the discovery of an unusual phase transition
in a ferroelectric dot for some critical value of the residual depolarizing field.

We thank P. Ghosez for discussion.  This work is supported by 
NSF grants DMR-0404335 and DMR-9983678, by ONR 
grants D 00014-01-1-0365, D 00014-04-1-0413 and D 00014-01-1-0600 and by DOE grant
DE-FG02-05ER46188.

\bibliography{lowdimdip10}

\begin{thebibliography}{19}
\expandafter\ifx\csname natexlab\endcsname\relax\def\natexlab#1{#1}\fi
\expandafter\ifx\csname bibnamefont\endcsname\relax
  \def\bibnamefont#1{#1}\fi
\expandafter\ifx\csname bibfnamefont\endcsname\relax
  \def\bibfnamefont#1{#1}\fi
\expandafter\ifx\csname citenamefont\endcsname\relax
  \def\citenamefont#1{#1}\fi
\expandafter\ifx\csname url\endcsname\relax
  \def\url#1{\texttt{#1}}\fi
\expandafter\ifx\csname urlprefix\endcsname\relax\def\urlprefix{URL }\fi
\providecommand{\bibinfo}[2]{#2}
\providecommand{\eprint}[2][]{\url{#2}}

\bibitem[{\citenamefont{Scott and de~Araujo}(1989)}]{scott}
\bibinfo{author}{\bibfnamefont{J.~F.} \bibnamefont{Scott}} \bibnamefont{and}
  \bibinfo{author}{\bibfnamefont{C.~A.~P.} \bibnamefont{de~Araujo}},
  \bibinfo{journal}{Science} \textbf{\bibinfo{volume}{246}},
  \bibinfo{pages}{1400} (\bibinfo{year}{1989}).

\bibitem[{\citenamefont{Fong et~al.}(2004)\citenamefont{Fong, Stephenson,
  Streiffer, Eastman, Auciello, Fuoss, and Thompson}}]{fong}
\bibinfo{author}{\bibfnamefont{D.}~\bibnamefont{Fong}},
  \bibinfo{author}{\bibfnamefont{G.}~\bibnamefont{Stephenson}},
  \bibinfo{author}{\bibfnamefont{S.}~\bibnamefont{Streiffer}},
  \bibinfo{author}{\bibfnamefont{J.}~\bibnamefont{Eastman}},
  \bibinfo{author}{\bibfnamefont{O.}~\bibnamefont{Auciello}},
  \bibinfo{author}{\bibfnamefont{P.}~\bibnamefont{Fuoss}}, \bibnamefont{and}
  \bibinfo{author}{\bibfnamefont{C.}~\bibnamefont{Thompson}},
  \bibinfo{journal}{Science} \textbf{\bibinfo{volume}{304}},
  \bibinfo{pages}{1650} (\bibinfo{year}{2004}).

\bibitem[{\citenamefont{Junquera and Ghosez}(2004)}]{junquera}
\bibinfo{author}{\bibfnamefont{J.}~\bibnamefont{Junquera}} \bibnamefont{and}
  \bibinfo{author}{\bibfnamefont{P.}~\bibnamefont{Ghosez}},
  \bibinfo{journal}{Nature} \textbf{\bibinfo{volume}{422}},
  \bibinfo{pages}{506} (\bibinfo{year}{2004}).

\bibitem[{\citenamefont{Kornev et~al.}(2004)\citenamefont{Kornev, Fu, and
  Bellaiche}}]{kornev}
\bibinfo{author}{\bibfnamefont{I.}~\bibnamefont{Kornev}},
  \bibinfo{author}{\bibfnamefont{H.}~\bibnamefont{Fu}}, \bibnamefont{and}
  \bibinfo{author}{\bibfnamefont{L.}~\bibnamefont{Bellaiche}},
  \bibinfo{journal}{Phys.\ Rev. Lett.} \textbf{\bibinfo{volume}{93}},
  \bibinfo{pages}{196104} (\bibinfo{year}{2004}).

\bibitem[{\citenamefont{Fu and Bellaiche}(2003)}]{fu}
\bibinfo{author}{\bibfnamefont{H.}~\bibnamefont{Fu}} \bibnamefont{and}
  \bibinfo{author}{\bibfnamefont{L.}~\bibnamefont{Bellaiche}},
  \bibinfo{journal}{Phys.\ Rev. Lett.} \textbf{\bibinfo{volume}{91}},
  \bibinfo{pages}{257601} (\bibinfo{year}{2003}).

\bibitem[{\citenamefont{Naumov et~al.}(2004)\citenamefont{Naumov, Bellaiche,
  and Fu}}]{naumov}
\bibinfo{author}{\bibfnamefont{I.~I.} \bibnamefont{Naumov}},
  \bibinfo{author}{\bibfnamefont{L.}~\bibnamefont{Bellaiche}},
  \bibnamefont{and} \bibinfo{author}{\bibfnamefont{H.}~\bibnamefont{Fu}},
  \bibinfo{journal}{Nature} \textbf{\bibinfo{volume}{432}},
  \bibinfo{pages}{737} (\bibinfo{year}{2004}).

\bibitem[{\citenamefont{Zhong et~al.}(1995)\citenamefont{Zhong, Vanderbilt, and
  Rabe}}]{zhong}
\bibinfo{author}{\bibfnamefont{W.}~\bibnamefont{Zhong}},
  \bibinfo{author}{\bibfnamefont{D.}~\bibnamefont{Vanderbilt}},
  \bibnamefont{and} \bibinfo{author}{\bibfnamefont{K.}~\bibnamefont{Rabe}},
  \bibinfo{journal}{Phys.\ Rev. B} \textbf{\bibinfo{volume}{52}},
  \bibinfo{pages}{6301} (\bibinfo{year}{1995}).

\bibitem[{\citenamefont{Kittel}(1996)}]{kittel}
\bibinfo{author}{\bibfnamefont{C.}~\bibnamefont{Kittel}},
  \emph{\bibinfo{title}{Introduction to Solid State Physics}}
  (\bibinfo{publisher}{John Wiley and Sons, Inc.}, \bibinfo{year}{1996}), pp.
  \bibinfo{pages}{386--388}, \bibinfo{edition}{seventh} ed.

\bibitem[{\citenamefont{Meyer and Vanderbilt}(2001)}]{meyer}
\bibinfo{author}{\bibfnamefont{B.}~\bibnamefont{Meyer}} \bibnamefont{and}
  \bibinfo{author}{\bibfnamefont{D.}~\bibnamefont{Vanderbilt}},
  \bibinfo{journal}{Phys.\ Rev. B} \textbf{\bibinfo{volume}{63}},
  \bibinfo{pages}{205426} (\bibinfo{year}{2001}).

\bibitem[{\citenamefont{Jensen}(1997)}]{jensen}
\bibinfo{author}{\bibfnamefont{P.~J.} \bibnamefont{Jensen}},
  \bibinfo{journal}{Ann. Physik} \textbf{\bibinfo{volume}{6}},
  \bibinfo{pages}{317} (\bibinfo{year}{1997}).

\bibitem[{\citenamefont{Vedemenko et~al.}(2003)\citenamefont{Vedemenko,
  H.P.Oepen, and J.Kirschner}}]{vedemenko}
\bibinfo{author}{\bibfnamefont{E.}~\bibnamefont{Vedemenko}},
  \bibinfo{author}{\bibnamefont{H.P.Oepen}}, \bibnamefont{and}
  \bibinfo{author}{\bibnamefont{J.Kirschner}}, \bibinfo{journal}{J. Magnetism
  and Magn. Materials} \textbf{\bibinfo{volume}{256}}, \bibinfo{pages}{237}
  (\bibinfo{year}{2003}).

\bibitem[{\citenamefont{Draaisma and de~Jonge}(1988)}]{draaisma}
\bibinfo{author}{\bibfnamefont{H.}~\bibnamefont{Draaisma}} \bibnamefont{and}
  \bibinfo{author}{\bibfnamefont{W.}~\bibnamefont{de~Jonge}},
  \bibinfo{journal}{J. Appl. Phys.} \textbf{\bibinfo{volume}{64}},
  \bibinfo{pages}{3610} (\bibinfo{year}{1988}).

\bibitem[{\citenamefont{Kittel}(1946)}]{kittel2}
\bibinfo{author}{\bibfnamefont{C.}~\bibnamefont{Kittel}},
  \bibinfo{journal}{Phys.\ Rev.} \textbf{\bibinfo{volume}{70}},
  \bibinfo{pages}{965} (\bibinfo{year}{1946}).

\bibitem[{\citenamefont{Mitsui and Furuichi}(1953)}]{mitsui}
\bibinfo{author}{\bibfnamefont{T.}~\bibnamefont{Mitsui}} \bibnamefont{and}
  \bibinfo{author}{\bibfnamefont{H.}~\bibnamefont{Furuichi}},
  \bibinfo{journal}{Phys.\ Rev.} \textbf{\bibinfo{volume}{90}},
  \bibinfo{pages}{193} (\bibinfo{year}{1953}).

\bibitem[{\citenamefont{Kaplan and Gehring}(1993)}]{kaplan}
\bibinfo{author}{\bibfnamefont{B.}~\bibnamefont{Kaplan}} \bibnamefont{and}
  \bibinfo{author}{\bibfnamefont{G.}~\bibnamefont{Gehring}},
  \bibinfo{journal}{J. Magn. Magn. Mater.} \textbf{\bibinfo{volume}{128}},
  \bibinfo{pages}{111} (\bibinfo{year}{1993}).

\bibitem[{\citenamefont{Bellaiche et~al.}(2000)\citenamefont{Bellaiche, Garcia,
  and Vanderbilt}}]{bellaiche1}
\bibinfo{author}{\bibfnamefont{L.}~\bibnamefont{Bellaiche}},
  \bibinfo{author}{\bibfnamefont{A.}~\bibnamefont{Garcia}}, \bibnamefont{and}
  \bibinfo{author}{\bibfnamefont{D.}~\bibnamefont{Vanderbilt}},
  \bibinfo{journal}{Phys.\ Rev. Lett} \textbf{\bibinfo{volume}{84}},
  \bibinfo{pages}{5427} (\bibinfo{year}{2000}).

\bibitem[{\citenamefont{Rhee et~al.}(1989)\citenamefont{Rhee, Halley, Hautman,
  and Rahman}}]{rhee}
\bibinfo{author}{\bibfnamefont{Y.~J.} \bibnamefont{Rhee}},
  \bibinfo{author}{\bibfnamefont{J.}~\bibnamefont{Halley}},
  \bibinfo{author}{\bibfnamefont{J.}~\bibnamefont{Hautman}}, \bibnamefont{and}
  \bibinfo{author}{\bibfnamefont{A.}~\bibnamefont{Rahman}},
  \bibinfo{journal}{Phys.\ Rev. B} \textbf{\bibinfo{volume}{40}},
  \bibinfo{pages}{36} (\bibinfo{year}{1989}).

\bibitem[{\citenamefont{Brodka}(2002)}]{brodka1}
\bibinfo{author}{\bibfnamefont{A.}~\bibnamefont{Brodka}},
  \bibinfo{journal}{Chem. Phys. Lett.} \textbf{\bibinfo{volume}{363}},
  \bibinfo{pages}{604} (\bibinfo{year}{2002}).

\bibitem[{\citenamefont{Brodka and Grzybowski}(2002)}]{brodka2}
\bibinfo{author}{\bibfnamefont{A.}~\bibnamefont{Brodka}} \bibnamefont{and}
  \bibinfo{author}{\bibfnamefont{A.}~\bibnamefont{Grzybowski}},
  \bibinfo{journal}{J. Chem. Phys.} \textbf{\bibinfo{volume}{117}},
  \bibinfo{pages}{8208} (\bibinfo{year}{2002}).

\end{thebibliography}

\newpage

\end{document}